# Materials and Design Strategies of Fully 3D Printed Biodegradable Wireless Devices for Biomedical Applications

Ju-Yong Lee[1,16]*, Jooik Jeon[2,16], Joo-Hyeon Park[1,16], Se-Hun Kang[1], Yea-seol Park[1], Min-Sung Chae[1,3], Jieun Han[1], Kyung-Sub Kim[1], Jae-Hwan Lee[1], Sung-Geun Choi[1], Sun-Young Park[4], Young-Seo Kim[1], Yoon-Nam Kim[1], Seung-Min Lee[1], Myung-Kyun Choi[1], Jun Min Moon[1], Joon-Woo Kim[5], Seung-Kwon Seol[6,7], Jeonghyun Kim[5], Jahyun Koo[8,9], Ju-Young Kim[10], Woo-Byoung Kim[11], Kang-Sik Lee[3], Jung Keun Hyun[2,12,13]*, Seung-Kyun Kang[1,14,15]*

[1]Department of Materials Science and Engineering, Seoul National University, Seoul, 08826, Republic of Korea.

[2]Department of Nanobiomedical Science and BK21 NBM Global Research Center for Regenerative Medicine, Dankook University, Cheonan 31116, Republic of Korea.

[3]Biomedical Engineering Research Center, Asan Medical Center, Seoul, 05505, Republic of Korea.

[4]Materials Safety Technology Research Division, Korea Atomic Energy Research Institute(KAERI), Daejeon 34057, Republic of Korea

[5]Department of Electronics Convergence Engineering, Kwangwoon University Seoul 01897, Republic of Korea.

[6]Smart 3D Printing Research Team, Korea Electrotechnology Research Institute (KERI), Changwon 51543, Korea

[7]Electrical Functionality Material Engineering, University of Science and Technology (UST), Changwon 51543, Korea

[8]School of Biomedical Engineering, College of Health Science, Korea University, Seoul 02841, Republic of Korea.

[9]Interdisciplinary Program in Precision Public Health, Korea University, Seoul 02841, Republic of Korea.

[10]Department of Materials Science and Engineering UNIST (Ulsan National Institute of Science and Technology), Ulsan 44919, Republic of Korea.

[11]Department of Energy Engineering, Dankook University, Cheonan 31116, Republic of Korea

[12]Department of Rehabilitation Medicine, College of Medicine, Dankook University, Cheonan, 31116, Republic of Korea

[13]Institute of Tissue Regeneration Engineering (ITREN), Dankook University, Cheonan, 31116, Republic of Korea

[14]Research Institute of Advanced Materials (RIAM), Seoul National University, Seoul, 08826, Republic of Korea

[15]Nano Systems Institute SOFT Foundry, Seoul National University, Seoul, 08826, Republic of Korea

[16]These authors contributed equally: Ju-Yong Lee, Jooik Jeon, Joo-Hyeon Park.

e-mail : seoulboy@snu.ac.kr, rhhyun@dankook.ac.kr; kskg7227@snu.ac.kr

**Abstract**

Three-dimensional (3D) printing of bioelectronics offers a versatile platform for fabricating personalized and structurally integrated electronic systems within biological scaffolds. Biodegradable electronics, which naturally dissolve after their functional lifetime, minimize the long-term burden on both patients and healthcare providers by eliminating the need for surgical retrieval. In this study, we developed a library of 3D-printable, biodegradable electronic inks encompassing conductors, semiconductors, dielectrics, thereby enabling the direct printing of fully functional, multi-material, customizable electronic systems in a single integrated process. Especially, conjugated molecules were introduced to improve charge mobility, energy level alignment in semiconducting inks. This ink platform supports the fabrication of passive/active components and physical/chemical sensors making it suitable for complex biomedical applications. Versatility of this system was demonstrated through two representative applications: (i) wireless pressure sensor embedded within biodegradable scaffolds, (ii) wireless electrical stimulators that retain programmable electrical functionality in vivo and degrade post-implantation. This work establishes a foundation of modules for autonomous, biodegradable bioelectronic systems fabricated entirely via 3D printing, with implications for personalized diagnostics, therapeutic interfaces, and transient medical devices.

## Introduction

Recent advances in sensors and Information and Communication Technologies (IoT) led to a new healthcare paradigm toward real-time and personalized precision medicine based on bio-interfaced electronic devices[1-2]. In particular, implantable medical devices have the advantages of direct and localized diagnosis and treatment of disease areas[3]. These devices have been developed across a wide range of target organs or tissues, such as cardiac pacemakers[4-5], brain implants for epilepsy[6], Parkinson's disease[7], cochlear implants[8], peripheral nerve stimulators[9-12] and continuous glucose monitoring system[13] for recording electrophysiological, mechanical, bio cues and therapeutic stimulation.

Biodegradable electronics have emerged as a transformative class of technologies for applications ranging from temporary biomedical implants and transient sensors to environmentally benign consumer devices.[14-16] Especially, biodegradable implantable devices not only eliminate the need for secondary removal surgery but also reduce long-term foreign body responses, lower infection risks, and synchronize therapeutic function with tissue healing.[17] Utilizing dissolvable metals/semiconductors and degradable polymers, researchers have demonstrated a broad array of functional systems including cardiac stimulator[18], patches[19], brain sensors[20-21], biomedical monitoring sensors[22-23], drug delivery system[24], and neurostimulator[25-27]. These devices have shown significant promise in diverse platforms, such as therapeutic implants in bioelectronic medicine[28], to sustainable wearable systems in green electronics[29], and adaptive interfaces in soft robotics.[30]

To date, most biodegradable electronic devices have been fabricated based on conventional microfabrication techniques, such as photolithography, thin-film deposition, etching, and soft lithography.[28, 31-32] While these methods offer high precision and device complexity, they often require cleanroom facilities, complex mask design, hazardous solvent-based processing, and multi-step alignment procedures.[33] These requirements can introduce compatibility challenges, particularly when working with biodegradable materials, which may have limited thermal or chemical stability.[33] Furthermore, integrating heterogeneous materials such as conductors, semiconductors, and insulators within a single device stack remains technically demanding when processing conditions vary significantly across layers.[34]

In contrast to top-down approaches, 3D printing enables bottom-up assembly of complex architectures with high material and design flexibility.[36] This additive technique provides several key advantages: (i) mask-free, rapid fabrication without the need for cleanroom infrastructure[37-38] (ii) seamless multi-material integration using customized printable inks[39] (iii) freedom to construct conformal, flexible, or volumetric geometries[40-42] (iv) easy customization and scalability, which are particularly advantageous for personalized medical devices.[43-44] Despite these advantages, the application of 3D printing to biodegradable electronics has remained largely limited to simple conductive traces[45] or recyclable conductive inks with non-degradable fillers[46-47]. This is due to the limited variety of materials[35, 48] compared to non-biodegradable materials, for which a wide range of advanced devices such as transistors[49], displays[50-52] and energy devices[53-56] have been extensively developed using 3D printing. Although biodegradable conductive inks and pastes have been investigated[57-59], realizing fully integrated, multifunctional devices including active electronic components and wireless modules through an entirely 3D-printed route remains a significant and largely unmet challenge.[60]

In this work, we present a comprehensive 3D-printing-based strategy for the fabrication of fully biodegradable, multifunctional electronic devices. By formulating printable, biodegradable inks for conducting, semiconducting, and insulating components, we demonstrate the direct fabrication of key functional units including Schottky diodes, p–n junction diodes, physical and chemical sensors, wireless pressure sensors, and wireless electrical stimulators using a unified, layer-by-layer additive manufacturing process. These devices exhibit pulsatile electrical stimulation, wireless functionality, and full biodegradability. Together, our results establish 3D printing as a versatile manufacturing platform for bespoke transient electronics, offering a pathway toward customizable, scalable, and biodegradable material-compatible device fabrication while achieving the level of functional complexity previously accessible only through conventional microfabrication.

## Main

**Figure 1** illustrates the development of fully 3D printable biodegradable electronic devices, starting from the design of functional bio-inks and progressing through the fabrication of electrical components via controlled ink deposition and stacking. These components, including passive/active components, physical/chemical sensors, were integrated into customized 3D device architectures. Among them, a wireless electrical stimulator was fabricated and successfully applied in animal experiments, demonstrating the translational feasibility of the platform. As shown in **Figure 1A**, four categories of biodegradable electronic inks were developed to serve as conductor, semiconductor, dielectric, and encapsulation/frame materials, and were co-printed using a multi-nozzle 3D printing system. To improve electrical conductivity, the conductive ink was subjected to electrochemical sintering via acid treatment, resulting in the formation of conducting bridges between printed structures.[61-62] For the semiconductive ink, an organic dye was employed to enhance charge transport by increasing the interconnectivity of semiconducting fillers and promoting junction formation.

Through three-dimensional configuration of the functional inks, both passive and active components were fabricated in various spatial arrangements, including lateral and vertically stacked orientations. These components were further integrated and encapsulated within application-specific 3D geometries, demonstrating a strategy for fully embedded packaging of structural electronics. As a representative demonstration, a cylindrical shaped wireless electrical stimulator was fabricated and inserted around the sciatic nerve in canine models. The device was wirelessly powered via RF coupling and successfully delivered monophasic electrical stimulation through its embedded leadless electrodes. **Figures 1B** and **1C** display scaffold-integrated versions of the wireless pressure sensor and the tubular shaped wireless electrical stimulator, respectively. **Figure 1D** shows an animal experiment demonstrating external wireless powering of the fully printed stimulator after implantation in a canine model, confirming the in vivo electrical functionality of the device.

**3D-printable biodegradable electronic inks**

**Figure 2** presents a comprehensive portfolio of 3D-printable biodegradable electronic inks, highlighting their electrical, rheological, and degradative characteristics. Each ink was formulated using a combination of inorganic functional fillers, biodegradable polymer matrices, and organic solvents. **Figure 2A** shows the conductivity of screen-printed traces composed of polycaprolactone (PCL) dissolved in tetrahydrofuran (THF) and zinc (Zn) microparticles (<10 μm). Electrochemical sintering using acetic acid[61] for ~10 minutes enabled significant conductivity enhancement. At ~40 vol% Zn, a conductivity of $\sim 2 \times 10^3$ S/m was achieved, which increased up to $\sim 10^5$ S/m at ~70 vol%. Sintering enhanced the conductivity by approximately $2 \times 10^7$-fold compared to the non-sintered trace for a 0.97 mm-high trace (**Figure S1**). These results can be modeled by the reactive diffusion model (**Supplementary Note 1**), wherein acetic acid permeates the printed structure, dissolves the $Zn(OH)_2$ oxide layer, and leads to the formation of $Zn(ac)_2$ passivation layer. Additionally, SEM imaging confirmed that the Zn fillers were interconnected after sintering, and XRD analysis revealed the formation of $Zn(ac)_2$ as a passivation layer (**Figure S2**). **Figure 2B** shows dielectric constant change by volume fraction difference of $Si_3N_4$ nanoparticles (~50 nm) blended in PCL/THF solution. Even a small increase in the volume fraction of dielectric nanoparticles (0 – 0.07) resulted in measurable enhancement of DC capacitance. This behavior was modeled using an interphase power-law framework for nanodielectrics (**Supplementary Note 2, Figure S3**). Modeling results indicate that the interphase volume fraction increases to approximately 0.6 when the $Si_3N_4$ filler volume fraction reaches 0.08 in the composite.

To ensure extrusion printability, tetraglycol (TG) was added as a lubricant/humectant to reduce filler–filler friction and minimize shrinkage during solvent evaporation. To preserve the conductive network close to the percolation threshold, the Zn/PCL ratio in the formulation was kept constant (~62 vol% for Zn), while varying the TG:THF ratio (1:2, 1:1, 2:1) to assess the influence of TG content. (**Figure 2C**) At a 1:2 ratio (TG 0.5 mL, THF 1 mL), the storage modulus saturated at ~$10^4$ Pa after ~5 minutes, indicating optimal rheological conditions for stable layer stacking. This effect was confirmed also for PBAT:CF ratio for Mo based inks (~29 vol%) (**Figure S4**).

Volume-fraction-dependent conductivity of several TG-added ink systems is summarized in **Figure 2D**. Zn-PCL exhibited conductivities up to $10^4$ S/m, whereas Mo-PBAT reached a maximum of ~307.5 S/m at approximately 30 vol%. In contrast, ZnO-PCL maintained a relatively low and nearly constant conductivity, limiting its potential for semiconductive applications. To address this limitation, biocompatible conjugated molecules, including brilliant yellow (BY), indigo (IND), and guanine (G), were introduced to facilitate charge transfer between ZnO particles. As shown in **Figure 2E**, the addition of BY and IND increased conductivity by factors of 31.8 and 2.87, respectively, relative to the pure ZnO ink.

This enhancement is hypothesized to arise from Fermi-level pinning between the ZnO surface and the molecular orbitals of the conjugated additives[63] While similar effects have been reported under high-pressure or high-temperature conditions in previous studies[64], the present results suggest that a comparable mechanism may operate under ambient processing conditions. Cross-sectional analysis of ZnO-PCL prints (**Figure 2F**) revealed densely packed ZnO domains interspersed with ZnO-free voids, likely resulting from solvent evaporation and possible micelle-like organization of tetraglycol during blending and drying. Furthermore, **Figure 2G** indicates that conjugated molecules with narrower direct band gaps produced greater conductivity enhancement. This observation suggests that molecules embedded between densely packed ZnO domains can improve charge transport, with the degree of enhancement correlating with the bandgap of the additive. Among the three molecules, BY demonstrated the most efficient electron conduction, likely due to favorable alignment of its energy levels with the Fermi level of ZnO nanoparticles. Overall, these results indicate that embedding appropriately selected conjugated molecules within ZnO-based inks can significantly enhance conductivity, while the proposed Fermi-level pinning mechanism provides a plausible explanation consistent with observed trends.

The rheological properties of the inks were characterized via shear stress–modulus measurements, all exhibiting yielding behavior as shown in **Figure 2H** ($ZnO_{BY}$-PCL ink is characterized in **Figure S5)**. **Figure 2I** displays a stacked, cylindrical lattice structure with representative inks (Zn-PCL, Mo-PBAT, ZnO-PCL, $Si_3N_4$-PCL) with a feature width of 0.4 mm, printed using a commercially available pneumatic extrusion 3D printer (BIOX™, Cellink Inc., Sweden). Further miniaturization to ~0.1 mm linewidths was achieved using a customized high-pressure dispensing printer (KERI, Korea), enabling fine-resolution patterning for resistors or semiconducting channel. For example, printed Zn-PCL resistors with a feature size of ~0.2 mm, length ~110 mm exhibited resistance values of approximately 30 Ω after sintering, demonstrating successful fabrication (**Figure S6**). Additionally, it was observed that inks printed perpendicular to the tensile direction exhibited lower moduli and elongation compared to those printed parallel to the tensile direction (**Figure S7**).

Encapsulation is essential in biodegradable electronics to prevent premature dissolution of water-soluble circuits.[32] A bilayer hydrophobic candelilla wax (CWAX) and relatively soft PBAT, which are biodegradable[65-66], was applied via 3D printing to provide water permeation resistance (**Figure 2J, Figure S8**). **Figure 2K** shows the resistance change of Zn-PCL traces (thickness ~400 μm) encapsulated with PBAT-only, PBAT–CWAX single layers, and PBAT–CWAX bilayers in PBS at 37 °C. The PBAT-only encapsulation doubled in resistance within 13 hours, while PBAT–CWAX reached 46 hours. The bilayer encapsulation further extended this to 150 hours, demonstrating significant improvement in device longevity through printed encapsulation strategies.

**3D-printed biodegradable passive and active circuit elements**

These inks and printing processes enable the all-in-one fabrication of 3D-printable, biodegradable passive and active electronic components with spatially programmable architectures. For passive components, Zn-PCL, $Si_3N_4$-PCL, and PBAT-based inks used as the electrode, dielectric, and structural layers, respectively, to construct capacitors and inductors with tunable 3D geometries and performance.

In **Figure 3A**, six micro-LEDs were successfully operated under a threshold voltage of 1.7 V using printed Zn-PCL traces, confirming sufficient charge transport capacity. **Figure 3B** demonstrates the integration of a flat Zn-PCL coil with a surface-mounted (SMD) capacitor and NFC chip, enabling wireless operation. The hybrid printed NFC device (**Figure S9**) exhibited resonance with a smartphone at ~10 MHz, as confirmed by return loss ($S_{11}$) measurements, and its operation was validated via NFC reader (**Movie S1**). To fabricate a vertically stacked 3D coil, a compact square-stacked configuration was implemented. Biodegradable magnetic composite ink composed of $Fe_3O_4$ and PCL was formulated and printed within the coil structure to enhance magnetic flux density (**Figure S10**). This magnetic layer was co-printed with Zn-PCL and PBAT to form 3D stacked inductor. Optical image and X-ray imaging shows the internal coil structure containing $Fe_3O_4$ -PCL layers (**Figure S11**). When connected to an LED and exposed to alternating magnetic fields (39.2 MHz, 460 mVpp), the 3D coil enabled wireless powering, resulting in visible LED activation (Figure 3C). **Figure 3D** shows a printed capacitor composed of a $Si_3N_4$-PCL dielectric layer sandwiched between Zn-PCL or Mo-PBAT electrodes. The dielectric composite displayed frequency-dependent behavior, with capacitance decreasing from ~70 pF at 1 kHz to ~10 pF at 100 kHz (**Figure 3E**), consistent with dielectric relaxation mechanisms such as dipolar and interfacial polarization suppression.

For active components, semiconducting ZnO-PCL inks blended with conjugated molecules, conductive inks (Zn-PCL, Mo-PBAT), and solid-state electrolyte (NaCl-agarose gel) were used to fabricate stackable diodes and transistors. Work functions of Zn-PCL and Mo-PBAT were measured using a Kelvin probe with a HOPG tip and determined to be 3.49 eV and 4.47 eV, respectively. The electronegativity of ZnO-PCL was calculated to be ~4.01 eV based on UPS, UV/VIS, and valence band edge measurements (**Figure S12**). Asymmetric junctions (ZnO/Zn: $\varphi = -0.51$ eV; ZnO/Mo: $\varphi = 0.47$ eV) formed effective Schottky barriers, resulting in diodes with on/off ratios of ~12 and high reproducibility (n = 10), as shown in **Figure 3F**. **Figure 3G** is 3D printed Schottky diode Zn-PCL / $ZnO_{BY}$-PCL / Mo-PBAT where similar rectification behavior was observed when BY was blended into the ZnO-PCL ink with higher currents (**Figures 3H**).

Additionally, the formation of p–n junctions was confirmed at the interface between 3D printable p-type Si-PCL (characterized in **Figure S13**) and ZnO-PCL in the presence of conjugated molecules, which further validating the utility of conjugated molecule for junction formation as well as conductivity enhancement. Without the conjugated molecule, a symmetric I–V response was observed, indicating the absence of a depletion zone (**Figure S14**). However, with the addition of a conjugated molecule, the diode exhibited an asymmetric I–V curve with an on/off ratio of ~33 (–10 V to 10 V) (**Figure 3I**) and avalanche breakdown at –20 V, confirming the formation of a depletion region. Using this junction, an N type/P type/N type - based transistor was constructed with a ZnO-PCL / p-type Si-PCL / ZnO-PCL configuration with conjugated molecule (**Figure 3J**, guanine; **Figure S15**, Indigo). Applying a positive gate voltage to the central p-region modulated the junction barrier, suppressing charge transport across the NPN path via junction potential modulation. **Figures 3K** presents an electrolyte-gated field effect transistor (Electrolyte-FET) composed of a ZnO-PCL channel, NaCl-agarose gel gate dielectric (1mM NaCl, 2% Agaorse), and Mo-PBAT gate electrode. The solid-state electrolyte allowed thickness-independent gating[67], despite the printed feature size being ~0.4 mm (**Figure S16**) beyond the limit of nanometer-thickness gate dielectric. The transistor demonstrated a mobility of 0.533 $cm^2/V \cdot s$ and an on/off ratio of $1.68 \times 10^4$ (–4 V to 4 V), as calculated from the saturation regime of the EGT transfer curve (**Figure 3L**).

**3D-printed biodegradable physical, chemical sensors and its structural adaptability**

A diverse set of biodegradable and 3D-printable sensors has been developed by integrating the biodegradable electronic inks shown in **Figure 2** with additional optimized formulations, including aluminum-doped ZnO ($ZnO_{Al}$)-PCL and $MoO_3$-PCL (**Figure S17**), enabling applications across both physical and chemical sensing modalities. In **Figure 4A**, resistive-type temperature sensor fabricated using Zn-PCL exhibited a negative temperature coefficient of resistance (TCR) of approximately 0.002 $K^{-1}$, with a highly linear response across the tested range ($R^2 = 0.998$). A UV-responsive sensor was developed using Mo-PBAT and $ZnO_{Al}$-PCL based inks (**Figure 4B**), showing consistent and reversible conductance changes upon repeated UV irradiation cycles (**Figure 4C**). A capacitive-type pressure sensor was constructed using a trench-structured Mo-PBAT/air/Mo-PBAT configuration (**Figure 4D, 4E**) owing to melt PBAT's self-supporting and spanning feature across gaps. Printed trench structured sensor showed repetitive capacitance change by the external loads (**Figure 4F**). pH sensor was printed using a $ZnO_{Al}$-PCL/Zn-PCL configuration in a two-electrode setup (referenced to Ag/AgCl) (**Figure 4G, 4H**), demonstrating a linear open circuit potential (OCP) shift with a sensitivity of –0.0316 V/pH within the pH range of 4–10 (**Figure 4I).** A glucose sensor was fabricated as a three-electrode system consisting of a $MoO_3$-ferrocene-glucosidase-PCL enzyme layer and a $MoO_3$-ferrocene-PCL mediating layer. This device exhibited a sensitivity of 0.345 μA/log(mM) within a concentration range of 0–20 mM (**Figure 4J, 4K**, and **4L**).

**Figure 5** highlights the structural versatility and integration capability of multi-material 3D-printed sensors, demonstrating both embedded and surface-mounted formats. Two primary strategies were used: (1) co-printing electronic elements within scaffold structures, and (2) direct deposition of sensors onto complex, static, or transformable surfaces. Embedding sensors directly into 3D-printed scaffolds can provide real-time diagnostics in biomedical implants.[68-69] Wireless pressure sensing was achieved using LC circuit-based resonant frequency shift. The system consisted of a trench-structured pressure capacitor (Mo-PBAT/air/Mo-PBAT) and a printed wireless antenna (Mo-PBAT/$Fe_3O_4$-PCL) (**Figure 5A, 5B**). Variations in applied pressure altered the capacitance, leading to detectable resonance frequency shifts (**Figure 5C**, **Movie S2**). Sensitivity values reached –0.193 MHz/g around 610 MHz for free-standing sensors, whereas scaffold-embedded configurations showed lower sensitivity (~0.0135 MHz/g) (**Figure 5D**). **Figures 5E–5G** show a 2 × 2 array of resistive temperature sensors printed on a stair-like static slope structure. Upon brief finger contact with a selected sensor, a localized temperature increase was confirmed via IR thermography, showing a distinct temperature change (~7 °C) relative to the surroundings. The sensor's resistance exhibited a sharp increase followed by a gradual recovery over time. **Figure 5H** illustrates an adaptive system in which two UV sensors were placed on opposite sides of a shape-shifting, auxetic-patterned PLA dome. Shape transformation was triggered by external heating (~80 °C), exploiting PLA's shape memory behavior (**Movie S3**). As shown in **Figure 5I**, the UV sensor facing the light source (3 mm distance) responded with a conductance increase from 2.37 nG to 19.4 nG upon dome expansion, while the sensor on the far and tilted side (4.5 mm) exhibited a smaller conductance shift from 2.97 nG to 4.64 nG. These results demonstrate that the combination of functional biodegradable inks and stereographic 3D-printing strategies enables the development of fully integrated sensor systems with diverse form factors and multimodal sensing capabilities. The adaptability to conformal or embedded structures makes these systems highly suitable for biomedical and environmental monitoring devices.

**3D-printed biodegradable wireless electrical stimulator and in vivo experiments**

**Figure 6A** shows the component sets for 3D-printable biodegradable stereographically integrated, and fully bioresorbable wireless stimulator. Main components of wireless stimulator consist of inductor (Zn-PCL/$Fe_3O_4$-PCL), capacitor (Zn-PCL/$Si_3N_4$-PCL/Zn-PCL), Schottky diode (Zn-PCL/$ZnO_{BY}$-PCL/Mo-PBAT), electrodes (Mo-PBAT), frame (PBAT). The wireless receiver, which is the radio frequency (RF) antenna, comprises a 10-turn square solenoid inductor (Zn-PCL) with a high permeability layer ($Fe_3O_4$-PCL), which enhances density of magnetic flux at the core, and a vertically stacked interdigitated capacitor. It wirelessly generates AC current by inductive coupling with an external transmission coil. The Schottky diode rectifies the AC pulse to a monophasic stimulating current, and electrodes form an interface to the nerve within a polygonal tubular shape. Electrochemical sintering was partially done for inductors and wires and the encapsulation layer was coated after sintering which electrically isolates the device from surrounding tissue and protects the device from early dissolution via biofluids. The device designs were converted into G-code-based layer-by-layer printing paths using slicing software and co printed materials, as shown in **Figure S18, Movies S4**.

**Figure 6B** shows the photographic image of a wireless biodegradable stimulator (ID, inner diameter = 8.8 mm; 11.2 mm (W)×14.8 mm (H)×12.8 mm (L)), where the dimension of wireless stimulator components are inductor (L, 11.2 mm (W)×11.2 mm (H)×12.8 mm (L)), capacitor (C, 11.2 mm (W)×1.2 mm (H)×12.8 mm (L)), and Schottky diode (D, channel dimension of 0.4 mm (W)×1.2 mm (H)×11.2 mm (L)). **Figure 6C** and **6D** show the performance of the biodegradable 3D-printed wireless stimulator. **Figure 6C** shows the typical power transfer curve with ~10.5 MHz resonant frequency via inductive coupling. **Figure 6D** shows monophasic output (amplitude 1 V, frequency 20 Hz, duration 200 μs, 1 kΩ loaded) generated by the external transmission antenna (~600 mVpp at 15 mm coupling distance). Generated monophasic output is comparable to early reported therapeutic stimulation for peripheral nerve[70-71].

**Figure 6E-6J** shows an in vivo demonstration of a 3D-printed bioresorbable wireless stimulator for sciatic nerve injury in animal models regarding in vivo operation. First, small animal test was conducted with a 3D printed wireless stimulator. Since the device was larger than the rat's sciatic nerve, the device was connected to Mo wire (**Figure S19**) for an interface with nerve and fully implanted in the subcutaneous region (**Figure 6E, Figure S20**). The leg trembled, and the compound muscle action potential (CMAP) was evoked (**Figure 6F, Movie S5**) when external coil wirelessly powered stimulator with programmed stimulation condition (amplitude 600 mVpp, frequency 20 Hz, duration 200 μs). The large animal test was conducted

with a 6-month-old canine model. Transection of the sciatic nerve at an approximate distance of 20 mm distal to the femoral joint, succeeded by the insertion of a tube-like 3D-printed stimulator into the proximal nerve segment (**Figure S21**). The proximal and distal segments were then sutured together to establish an interface with the customized device on the injured nerve (**Figure 6G**). Additionally, C-arm imaging confirmed that the stimulator retained its structural integrity even when covered by muscle tissue (**Figure 6H**). The limb was positioned within an external coil, aligning the direction of the implanted solenoid-like inductor with the coil to maximize the magnetic flux (**Figure 6I**). Since CMAP did not exhibit clear signal changes (**Figure S22**), compound nerve action potential (CNAP) was measured instead, as it exhibits approximately fourfold greater amplitude in response to the same stimulus when recorded from the canine sciatic nerve (**Figure S23**). **Figure 6J** and **Movie S6** present the evoked CNAP recorded during wireless stimulation, with a peak amplitude of approximately 3 mV. The CNAP exhibits a biphasic response characterized by an initial depolarization followed by a gradual decay and subsequent re-excitation, with a temporal interval of 50 ms between successive peaks. This response was induced by wirelessly delivered electrical pulses generated by the external coil (1.1 Vpp, 20 Hz, 200 μs pulse width). This successful demonstration shows the ability of the bioresorbable 3D-printed device to induce neural activity in a large animal model.

**Biodegradability and biocompatibility of 3D printed biodegradable wireless stimulator**

A unique feature of this 3D-printed stimulator is that all constitution materials are biodegradable to eliminate the necessity of secondary surgery after regeneration treatment ends. Filler materials (Zn, Mo, ZnO, $Fe_3O_4$, $Si_3N_4$) dissolve into soluble ions[32, 72-73], hydroxides ($Zn(OH)_2$, $H_2MoO_4$, $Si(OH)_4$, $Fe(OH)_2$) [32, 72-73], gases ($H_2$, $NH_3$) [32, 72-73]. Matrix materials (PCL, PBAT) hydrolyzed into monomers and generate byproducts of caproic acid, succinic acid, butyric acid, valeric acid, terephthalic acid, and adipic acids and are excreted from the body by urine or feces[65, 74]. Conjugated molecules (e.g. brilliant yellow, indigo, etc.) directly absorb into bio-fluid or as a byproduct after phagocytosis by macrophage[75-79]. Detailed dissolution chemistries of the constitution materials are described in **Tables S1, S2**. **Figure 7A** shows the dissolution behavior of a 3D-printed stimulator (0.792 g) fitted on a Pasteur pipette in phosphate-buffered saline (PBS) with accelerated condition (pH 6.8 – 7.8, Lipase, 45 °C). The interface of electronic ink/frame layers and inorganic filler dissolved in ~3-weeks and 8-weeks, and all remaining materials vanished after ~14-weeks.

**Figure 7B** presents the in vivo degradation of the 3D-printed bioresorbable stimulator implanted at dorsal subcutaneous region of SD rats over an 18-weeks. The degradation process was monitored through X-ray imaging. In this study, the accelerated sample is also prepared along with a pristine sample to observe rapid degradation and to check the biocompatibility of byproduct (**Figure S24**). To create the accelerated samples, we placed them in gaseous chloroform and the absorbed solvents leaked out in ethanol to make them porous, allowing more water to penetrate the PBAT frame. The pristine stimulator showed slow degradation with its main body integrity preserved after 18-weeks. Conversely, in the accelerated sample, gradual degradation was observed in both the frame materials and electronic components by the 8-weeks, and substantial degradation had become evident by approximately the 18-weeks. Consequently, it is necessary to consider a method that can increase the surface area of electronics on demand to resorb faster, and recently, ultrasonic wave has been studied to remove biodegradable electronic devices after operation[25]. Likewise, fracture occurred when ultrasonic waves were applied to the 3D printed structural materials (PBAT) inside water and 3D printed stimulator inside subcutaneous region in rat model (**Figure S25**) confirming more possibility of rapid/on demand degradation.

Also, the biocompatibility of the decomposed materials was thoroughly assessed through blood tests conducted at 6-weeks intervals to monitor potential inflammation (**Figure 7C**). Hematological analysis of rats implanted with both pristine and accelerated stimulator shows overall health status and no significant differences in the biomarkers for immune response and cellular levels of the blood were observed, which matches the results from the non-implanted control rat. Additionally, consistent weight increment in both groups, which confirms no induction of fatal diseases, verifies biocompatibility of the stimulator and its byproducts.

**Conclusions**

This study presents the development of fully 3D-printable, biodegradable electronic devices through the design of functional bio-inks tailored for conductors, semiconductors, dielectrics, and encapsulation materials. Utilizing multi-nozzle 3D printing and material formulation and conductivity enhancement strategies, various passive and active components were fabricated with spatial programmability, enabling customizable three-dimensional architectures and seamless integration within complex structures. A diverse range of biodegradable sensors, including physical, chemical, and biosensing types, was integrated into conformal or scaffolded architectures. These systems demonstrated structural adaptability and real-time sensing capabilities suitable for biomedical implants and environmental monitoring. Incorporating wireless communication elements allows fully embedded, miniaturized sensor platforms with enhanced functional integration. A fully bioresorbable 3D-printed wireless electrical stimulator was fabricated and validated in vivo, delivering effective monophasic stimulation to peripheral nerves in animal models, highlighting the clinical potential of this device. Accelerated in vivo degradation studies showed full device dissolution within 14 to 18 weeks without adverse inflammatory or systemic effects, supported by comprehensive biocompatibility assessments. While challenges remain in enhancing printing resolution and mechanical softness, this work establishes a versatile and scalable approach to fabricating fully biodegradable wireless implantable bioelectronics. The combination of tailored materials, embedded multifunctional components, and wireless operation offers a promising platform for future development of intelligent 3D soft robotics, autonomous diagnostic implants, and therapeutic devices with minimal invasiveness and improved patient outcomes.

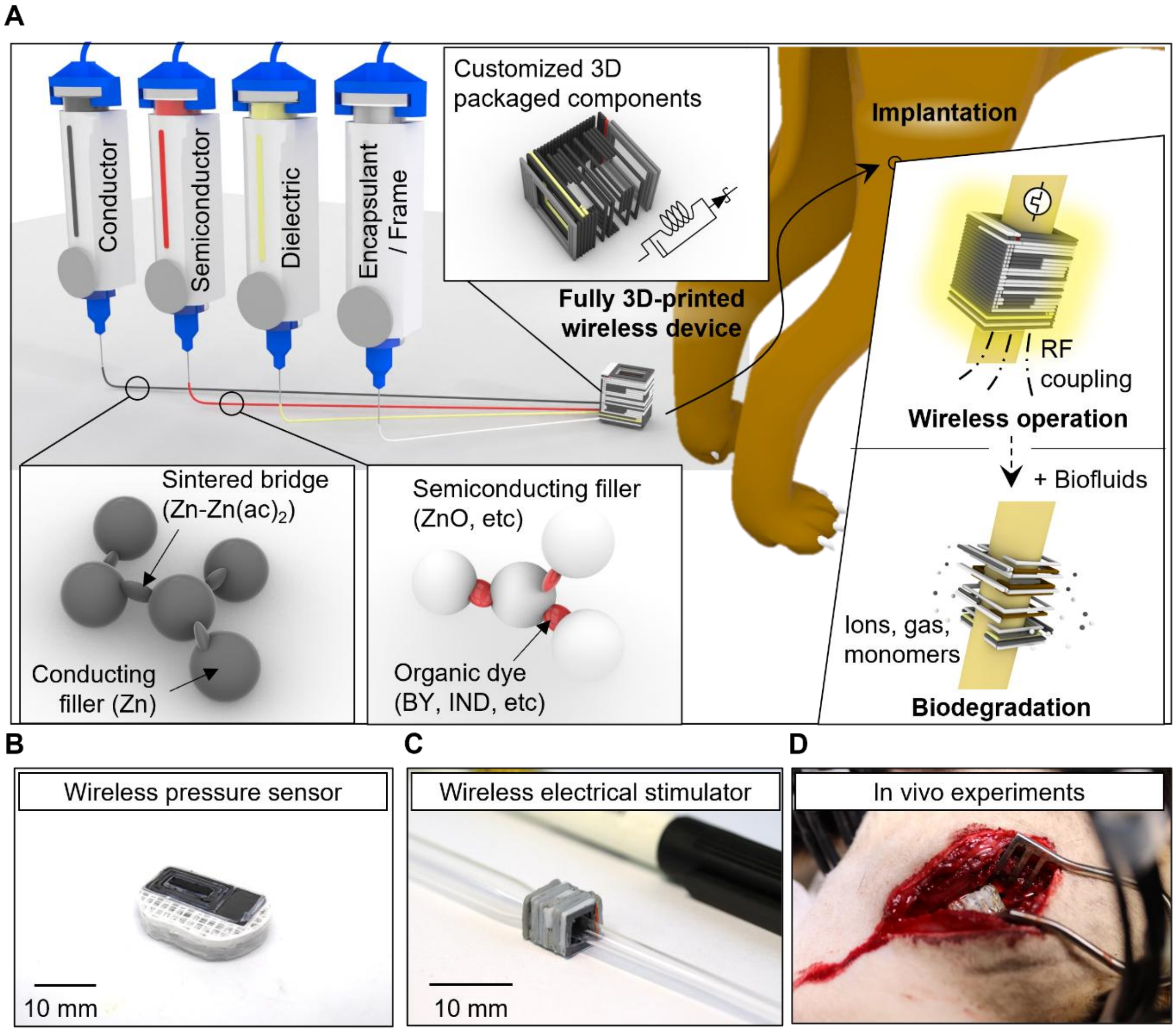

**Figure 1. Concept and Overview of the fully 3D printed biodegradable electronics for wireless implantable devices**

(A) Schematic illustration of conductivity enhancement strategy of conducting/semiconducting inks, design strategy for multi-material printed device, and transient operation of 3D printed wireless device. $Zn(ac)_2$ is Zinc acetate layer which is formed after electrochemical sintering in Zinc conducting ink. Organic dyes such as brilliant yellow (BY) and indigo (IND) are used for conductivity enhancement of ZnO or p-Si semiconducting ink. Here, the wireless stimulator including diode, inductor, capacitor, wires, and electrodes is built into the cylindrical structure to surround the nerve. After its wireless electrical stimulation, the stimulator can be degraded/absorbed by biofluids. (B) Image of 3D printed biodegradable wireless pressure sensor embedded in PCL scaffolds. (C) Image of 3D printed biodegradable cylindrical wireless electrical stimulator embracing Pasteur pipette (ID 5.6 mm). (D) Image of in vivo experiments for wireless electrical stimulation with 3D printed biodegradable device on sciatic nerve of canine model.

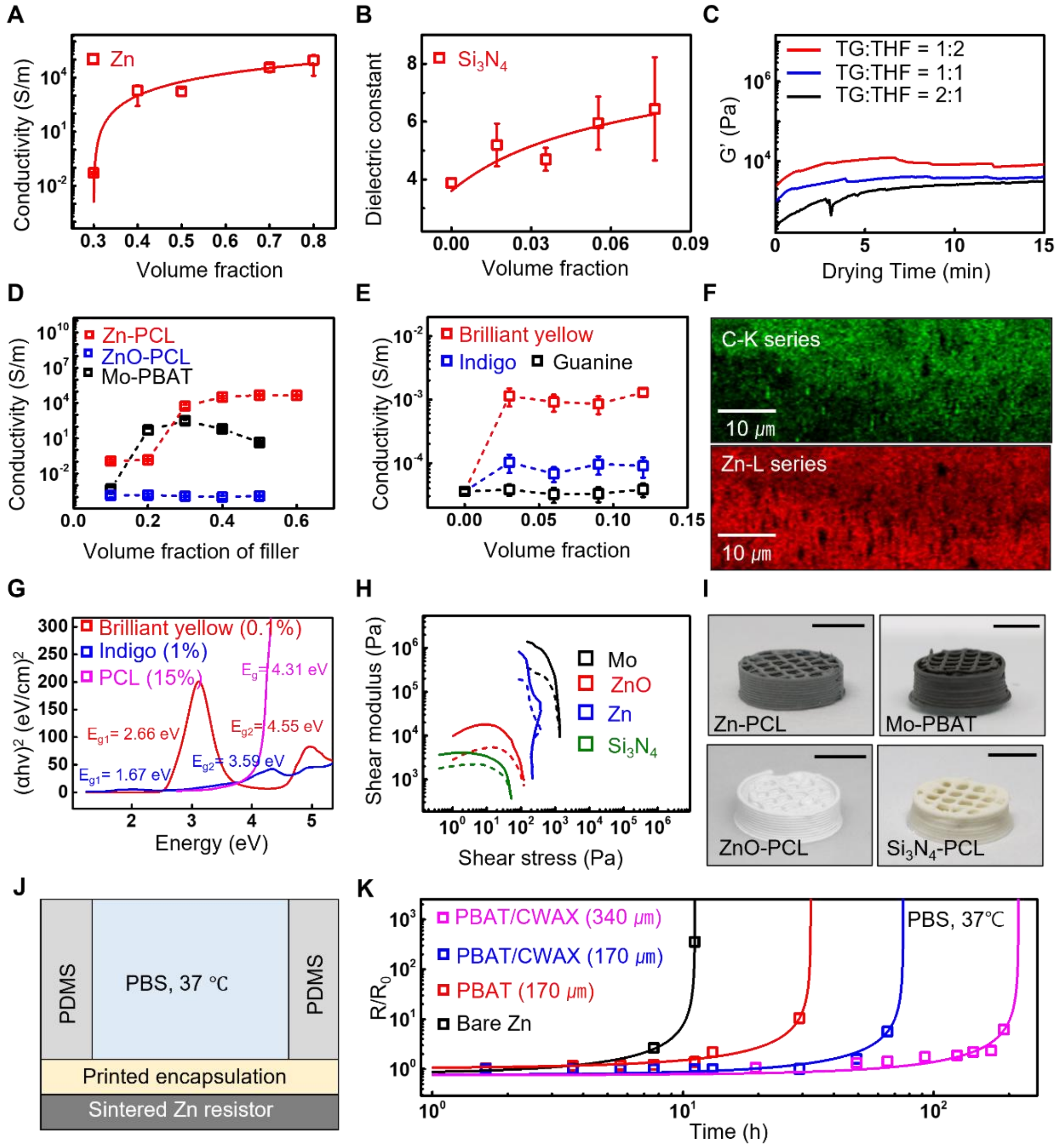

**Figure 2. Electrical, rheological, and dissolution properties of sets of biodegradable 3D-printable electronic inks.**

(A) Conductivity variation of screen-printed Zn-PCL(THF) ink by volume fraction of Zn fillers (experimental data, dot; Belehradek model fitting, line). (B) Dielectric constant of $Si_3N_4$-PCL(THF) film by volume fraction of $Si_3N_4$ fillers. (experimental data, dot; Interphase power model-based fitting, line) (C) Storage shear modulus change by drying time with different ratio between humectant (TG) and solvent (THF) in case of Zn-PCL ink. (D) Conductivity variation of Zn-PCL(THF), ZnO-PCL(THF), Mo-PBAT(CF) inks with TG by volume fraction of filler materials. Volume ratio of THF or CF : TG = 2:1. (E) Conductivity of ZnO-PCL(THF/TG) ink with addition of bioresorbable conjugated molecules (Guanine, black; Indigo, blue; Brilliant yellow, red). Volume fraction of ZnO is ~35%. (F) EDS images of sectioned surface of printed ZnO-PCL(THF/TG) ink (upper, C-K series; lower, Zn-L series). (G) Tauc-plot of Brilliant yellow 0.1% (w/v), Indigo 1% (w/v), and PCL 15% (w/v). (H) Shear modulus change by shear stress applied to Mo-PBAT(CF/TG), ZnO-PCL(THF/TG), Zn-PCL(THF/TG), and $Si_3N_4$-PCL(THF/TG) inks (Solid, storage modulus; dash, loss modulus). (I) Images of 3D printed cylindrical shape of Zn-PCL(THF/TG), Mo-PBAT(CF/TG), ZnO-PCL(THF/TG), and $Si_3N_4$-PCL(THF/TG) with grid pattern infill of 40%. Scale bar = 5 mm. (J) Schematics of dissolution behavior test platform for electrochemically sintered/3D printed Zn-PCL trace. (K) Dissolution behavior of 3D printed Zn-PCL traces (thickness 400 μm) measured using the variation of electrical resistance with various 3D-printed encapsulating layers (Bare Zn XX μm thick, black; PBAT 170 μm thick, red; PBAT-CWAX 170 μm thick, blue; PBAT-CWAX 340 μm thick, magenta) in PBS at 37 °C (experimental data, dot; theoretical fitting, line).

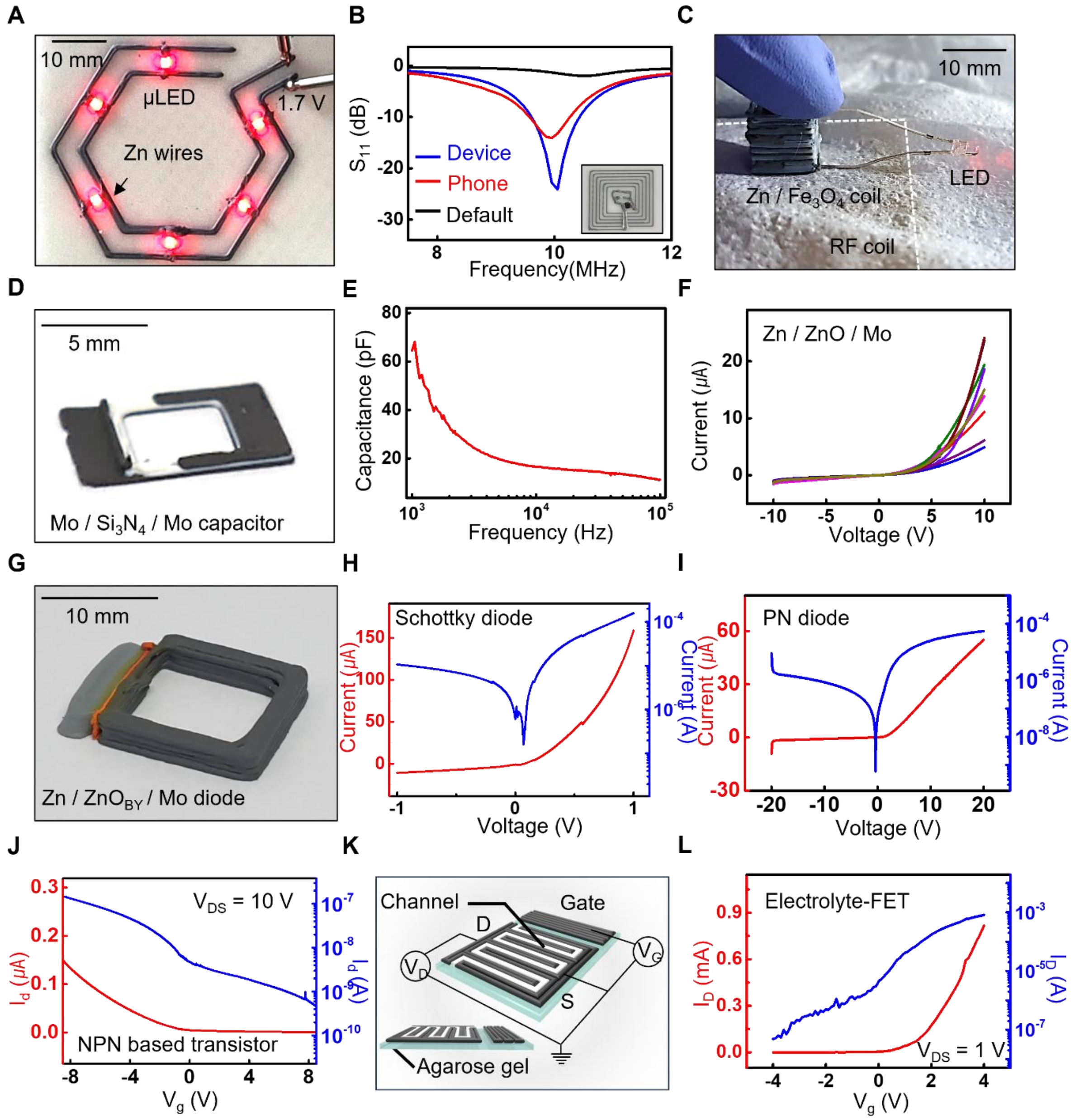

**Figure 3. Electrical characterization of sets of biodegradable 3D-printable passive and active circuit elements.**

(A) Image of 3D-printed Zn-PCL wires connected to six micro-LEDs and digital multi-meter (Voltage 1.7 V) (B) Return loss ($S_{11}$) by frequency for smartphone and hybrid printed NFC device combing 3D printed biodegradable flat coil (Zn-PCL on PLA substrate) and SMD chips (capacitor, Ntag123) (black, default; red, smart phone; blue, hybrid NFC device). Inset is smartphone interacting with hybrid NFC device. (C) Image of wirelessly powered LED connected to 3D printed square-stacked coil (conducting line, Zn-PCL; magnetic coupling agent, $Fe_3O_4$-PCL; frame, PBAT). Transmission coil is placed under kimwipe transmitting 460 mVpp, 39.2 MHz. (D) Image of 3D printed biodegradable capacitor with Z-axis interdigitated form. (E) Capacitance of 3D-printed Mo-PBAT/$Si_3N_4$-PCL/Mo-PBAT capacitor by frequency. (F) I-V curves of 3D printed Zn-PCL/ZnO-PCL/Mo-PBAT Schottky diode with 10 different samples from -10 V to 10 V. (G) Image of Zn-PCL/$ZnO_{BY}$-PCL/Mo-PBAT Schottky diode. (H) I-V curves of 3D printed Zn-PCL/$ZnO_{BY}$-PCL/Mo-PBAT Schottky diode. (I) I-V curves of 3D-printed p-type $Si_G$-PCL/$ZnO_G$-PCL from -20 V to 20 V. (J) I-V curves of 3D printed $ZnO_G$-PCL/p-type $Si_G$-PCL/$ZnO_G$-PCL transistor from -8 V to 8 V with source-drain voltage of 10 V. (K) Schematic diagram of electrolyte gated field effect transistor (Electrolyte-FET) (Electrode, Mo-PBAT; channel, ZnO-PCL; electrolyte, 1 mM NaCl agarose 2% gel). (L) Transfer curve of Electrolyte-FET with gate voltage from -4 V to 4 V with source-drain voltage of 1 V.

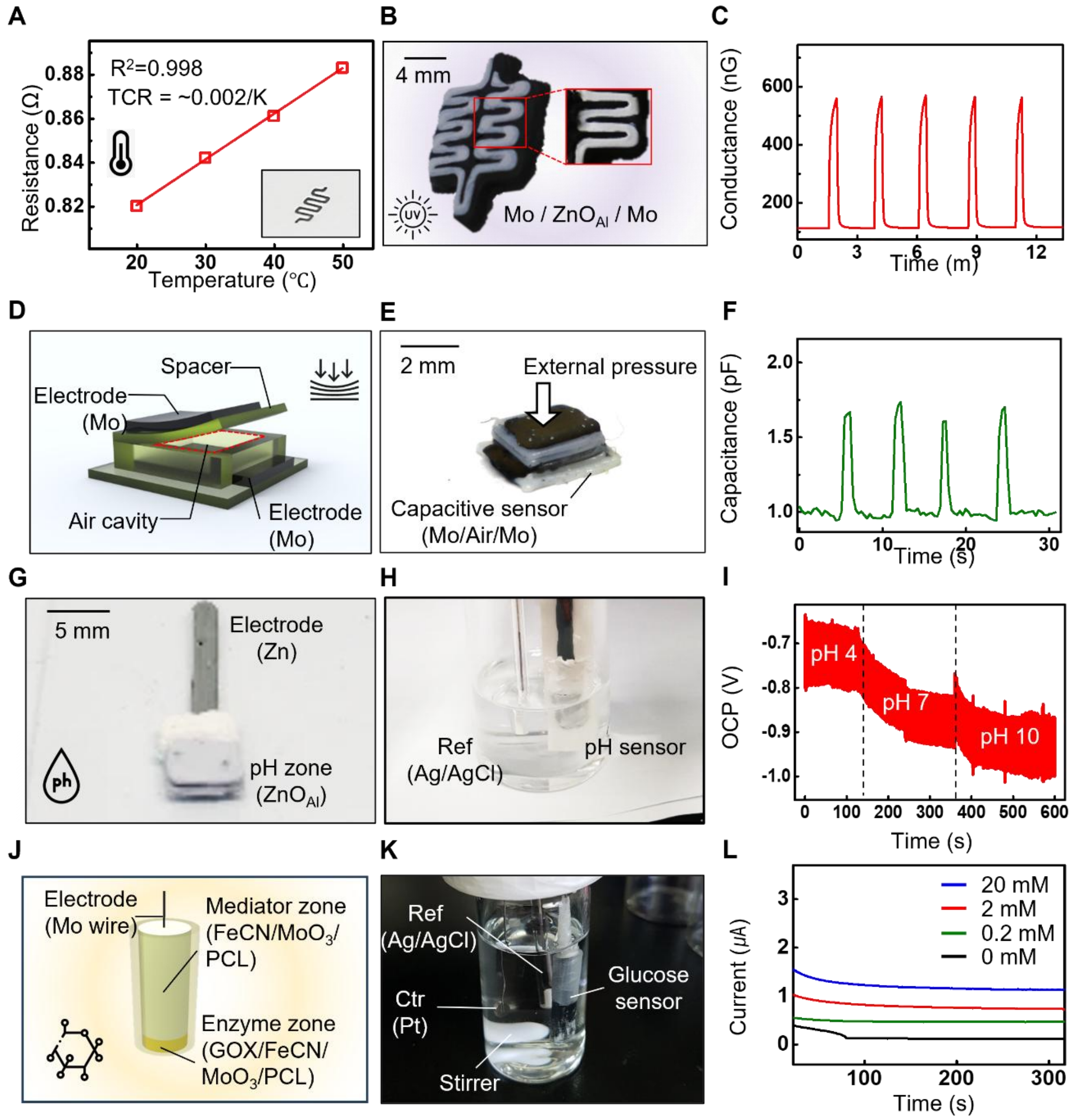

**Figure 4. Electrical characterization of 3D-printed biodegradable physical-, chemical-, bio-sensors.**

(A) Resistance changes of temperature sensor (Zn-PCL resistor) by temperature difference (20 ℃, 30 ℃, 40 ℃, 50 ℃). Inset is 3D printed biodegradable temperature sensor. (B) Image of 3D printed biodegradable UV sensor (Mo-PBAT electrode/ZnO$_{Al}$ channel/Mo-PBAT electrode). (C) Conductance changes of UV sensor by pulsatile UV irradiation. (D) Schematic illustration of pressure sensor (Mo-PBAT electrode/PBAT spacer/Air cavity/Mo-PBAT electrode). (E) Image of 3D printed biodegradable pressure sensor. (F) Capacitance changes of pressure sensor by repetitive pressure (~1 MPa). (G) Image of 3D printed biodegradable pH sensor (Zn-PCL electrode/$ZnO_{Al}$-PCL sensing zone). (H) Electrode system for pH sensor measurement set-up (reference, Ag/AgCl; working, pH sensor) (I) Open circuit potential (OCP) changes of pH sensor by pH difference (pH 4, pH 7, pH 10). (J) Schematic illustration of glucose sensor (GOX/FeCN/$MoO_3$-PCL, Enzyme zone; FeCN/$MoO_3$-PCL, Mediator zone; Mo wire, Electrode). (K) Three-electrode system for glucose sensor measurement set-up (reference, Ag/AgCl; working, glucose sensor; control, Pt) with -0.02 V applied. (L) Current change of Glucosidase-Ferrocene-$MoO_3$-PCL based glucose sensor by different glucose concentrations (0 mM, 0.2 mM, 2 mM, 20 mM).

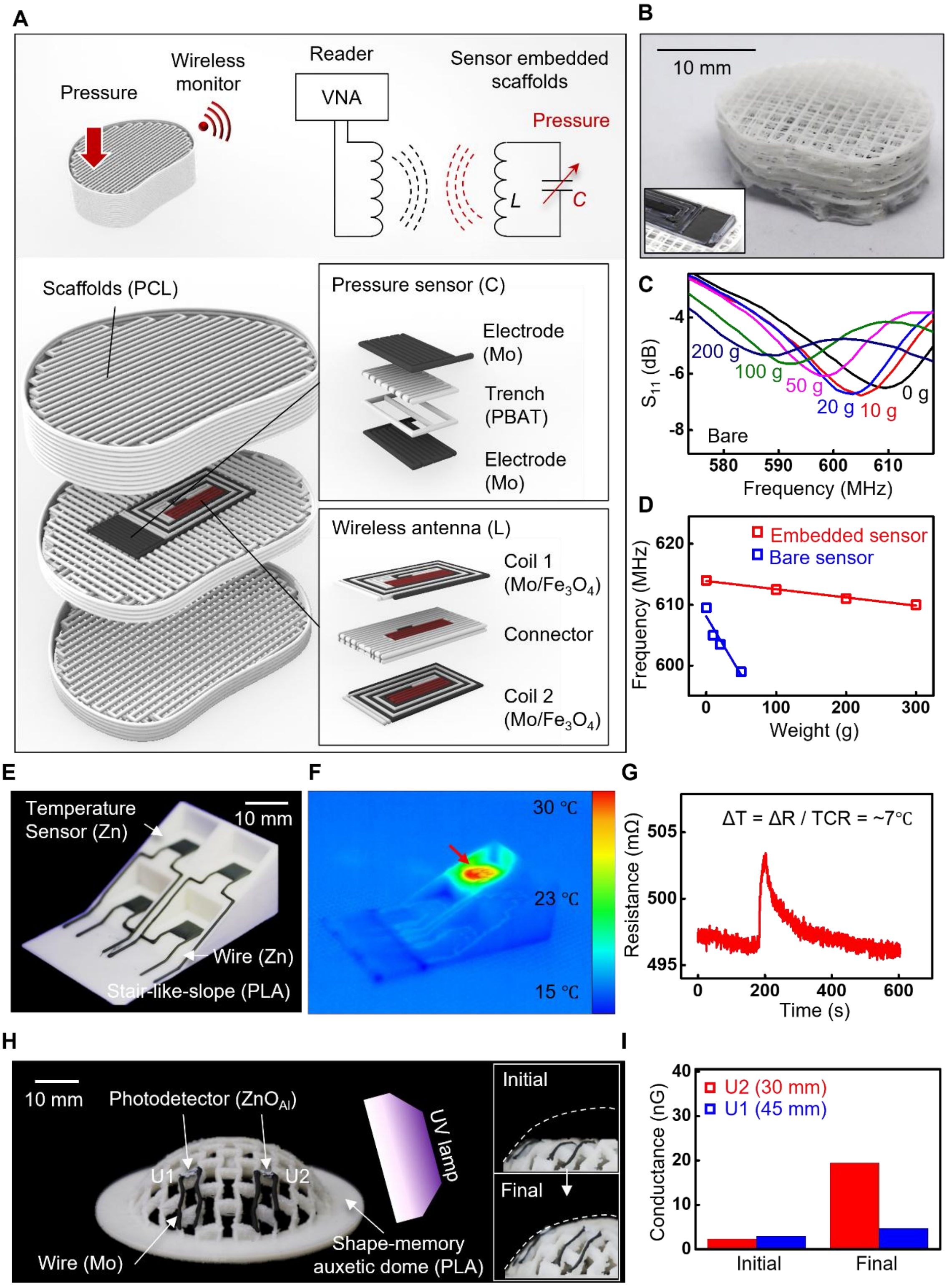

**Figure 5. Integration of sensors with various structure form factors (Embedded, on-static-surface, on-dynamic-surface).**

(A) Illustration of wireless pressure sensor embedded in PCL scaffolds consisting of a wireless antenna ($Fe_3O_4$-PCL/Mo-PBAT) and capacitive pressure sensor (Mo-PBAT/PBAT trench/Mo-PBAT), and wireless monitoring system consists of LC circuit/readout circuit diagram. (B) Image of 3D printed biodegradable wireless pressure sensor embedded PCL scaffolds. Inset is an image of internal wireless pressure sensor. (C) Resonance frequency changes of wireless pressure sensor without scaffolds by external loads (0 g, 10 g, 20 g, 50 g, 100 g, 200 g) on the sensor. (D) Resonance frequency changes of wireless pressure sensor with and without scaffolds. (E) Image of 3D-printed 2×2 temperature sensor (Zn-PCL) array on stair-like sloped structure (PLA). (F) IR image of 2×2 temperature sensor array where left/top part sensor was touched by finger. (G) Resistance changes of left/top part sensor when finger was contacted on/released from the sensor. (H) Overall images of UV sensors (Mo-PBAT/$ZnO_{Al}$-PBAT/Mo-PBAT) on PLA auxetic dome. U2 (30 mm) is closer to UV lamp than U1 (45 mm). Inset is serial images of shape-shifting process of UV sensors on PLA auxetic dome by the external heat transfer by heat gun (~80 °C). (I) Conductance changes of UV sensors (U1, U2) before and after shape shifting with constant UV lamp irradiation.

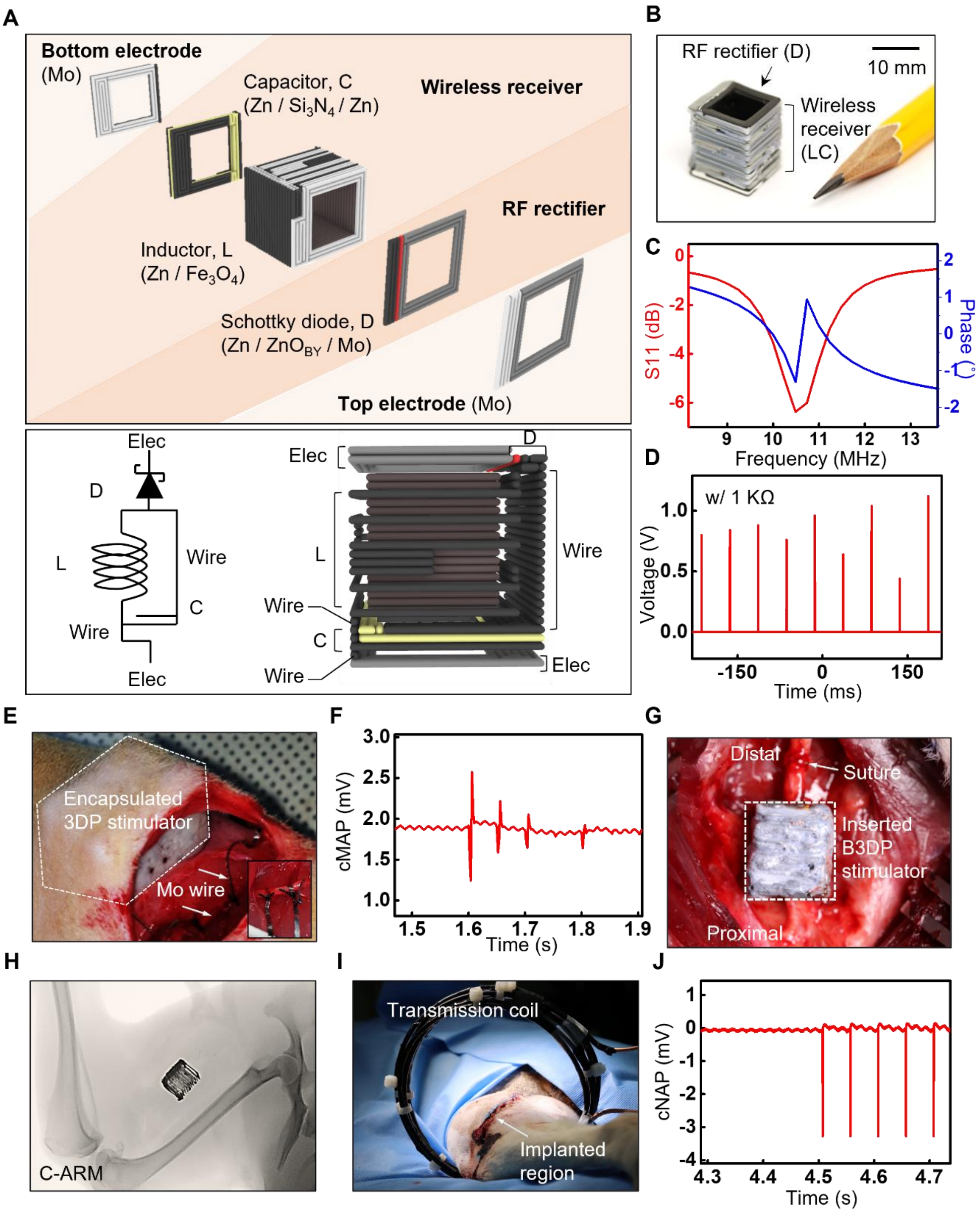

**Figure 6. Operation and acute in vivo experiment of a 3D packaged, biodegradable, and wireless stimulator for the sciatic nerve in rat and canine models.**

(A) Schematic illustration of the fully 3D-printed biodegradable wireless electrical stimulator in customized structure for injured peripheral nerve with an exploded view and translucent view of packaged electronic components with circuit diagram. Two electrodes (Mo-PBAT), wireless antenna including capacitor (Zn-PCL electrode/$Si_3N_4$-PCL dielectric/Zn-PCL electrode) and inductor (Zn-PCL coil / $Fe_3O_4$-PCL magnetic coupling agent), RF rectifier (Zn-PCL/$ZnO_{BY}$-PCL/Mo-PBAT Schottky diode). Each electronic component is interconnected with Zn-PCL wires and embedded in the nerve surrounding hollow cylindrical structure (PBAT) which could fit the peripheral nerve. (B) Image of a customized wireless stimulator with Wireless power receiver and RF rectifier (ID 8.8 mm). (C) RF behavior of the 3D-printed wireless stimulator (red, $S_{11}$; blue, phase) with ~10.5 MHz resonance frequency. (D) Generated monophasic pulse type output voltage (20 Hz, 200 μs) from 3D-printed wireless stimulator during external amplified input voltage of 600 mVpp applied at 1 – 1.5 cm distance with 1 kΩ load. (E) Image of implanted 3D-printed bioresorbable (B3DP) stimulator in subcutaneous region in rat model connected to sciatic nerve via Mo wires. Inset is showing nerve interfacing with Mo wires. (F) CMAP measurements with wireless stimulation (transmission voltage ~600 mVpp, 10 - 25 MHz, pulse 200 μs, interval 50 ms) in the rat model. (G) Image is inserted B3DP wireless stimulator embracing sciatic nerve of canine model. (H) C-ARM image of implanted B3DP wireless stimulator in canine leg. (I) Image of wireless powering system for implanted B3DP wireless stimulator using with transmission coil (J) CNAP measurement with wireless stimulation (transmission voltage ~1.1 Vpp, 10 - 25 MHz, pulse 200 μs, interval 50 ms) in the canine model.

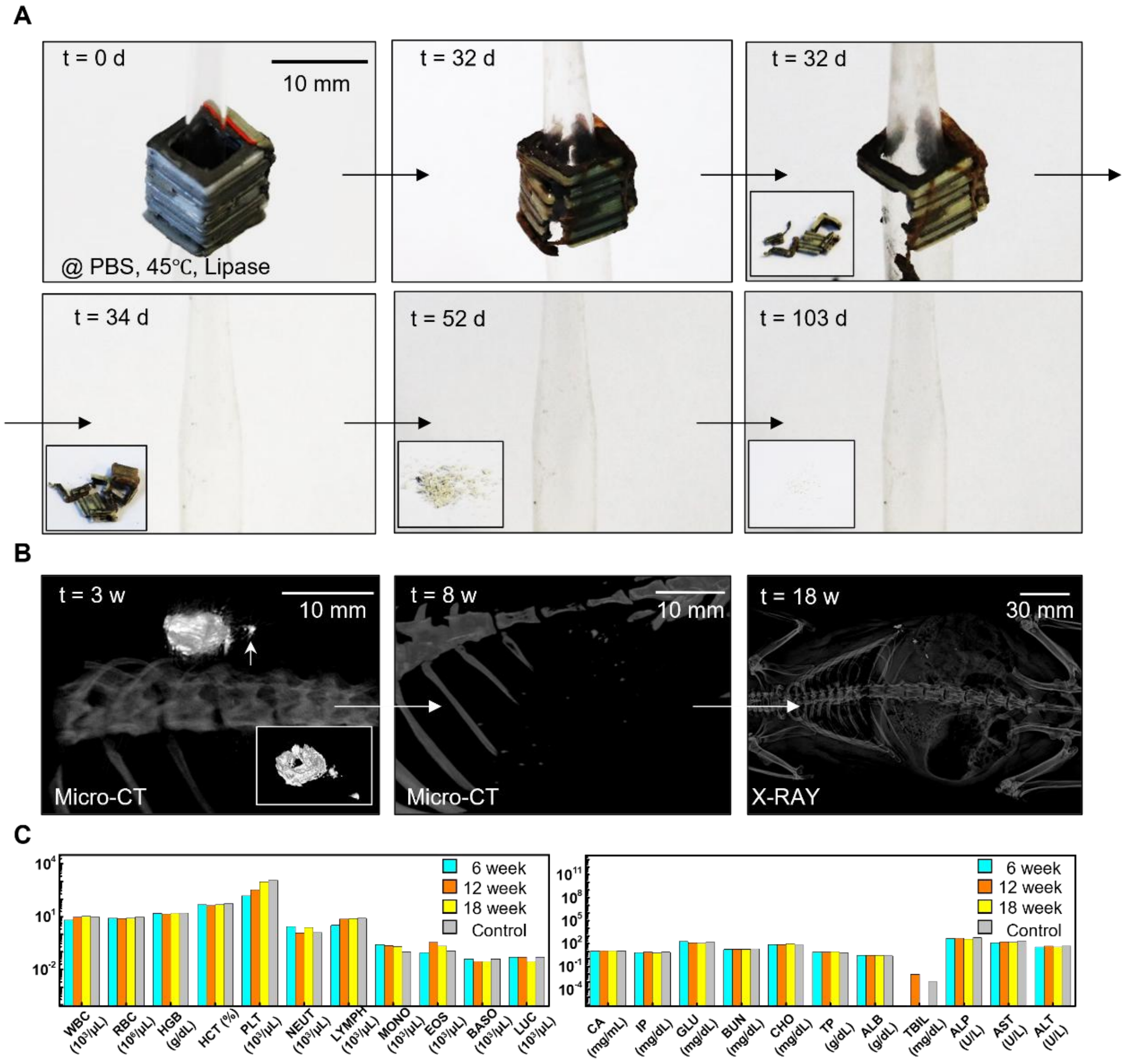

**Figure 7. In vitro and In vivo degradation behavior of 3D printed wireless stimulator**

(A) Serial images of in vitro degradation of 3D printed biodegradable stimulator in PBS, 45 °C, lipase condition for 103 days. Inset images show degradation behavior of the parts detached from the stimulator inserted to Pasteur pipette. (B) Serial Micro-CT and X-ray images for accelerated in vivo degradation of 3D-printed biodegradable stimulator in rat model for 18-weeks. Inset image in 3-weeks is a perspective view for the stimulator in vivo. (C) Hematological analysis of a rat implanted with a 3D printed biodegradable stimulator. Acronyms at blood analysis followed as: WBC (White Blood Cell), RBC (Red Blood Cell), HGB (Hemoglobin), HCT (Hematocrit), PLT (Platelet), NEUT (Neutrophil), LYMPH (Lymphocyte), MONO (Monocyte), EOS (Eosinophil), BASO (Basophils), LUC (Large

Unstained Cells); Ca (Calcium), IP (Inorganic Phosphorus), Glu (Glucose), BUN (Blood Urea Nitrogen), CHO (Cholesterol), TP (Total Protein), ALB (Albumin), TBIL (Total Bilirubin), ALP (Alkaline Phosphatase), AST (Aspartate Aminotransferase), and ALT (Alanine Aminotransferase).

**Methods**

*Materials used for biodegradable 3D-printable electronic inks and devices*

Filler: Zinc powder (Zn, dust, <10 µm, 98%, 209988-1KG, Sigma-Aldrich, Inc), Molybdenum powder (Mo, 800 nm, 99.9%, US Research Nanomaterials, Inc), Tungsten powder (W, 5 µm, 99.9 %, US Research Nanomaterials, Inc), Iron powder (Fe, 800 nm, 99.9 %, US Research Nanomaterials, Inc), Zinc Oxide (ZnO, 99.9+ %, 500 nm, US Research Nanomaterials, Inc), Zinc Oxide doped with 2 wt% Aluminum powder ($ZnO_{Al}$, 99.9 %, 300 nm, US Research Nanomaterials, Inc), Milled p-type silicon wafer (4” x 0.525 mm, <100>, B-doped, 0.001-0.002 ohm·cm, prime, Taewon scientific Co., Ltd) 100 - 300 nm, in pH 11 with DLS analysis (Milling condition, 600 RPM, 12 h, 12 g wafer with Zirconia ball 10mm : 3mm = 70 g : 30 g), Magnesium Oxide powder (MgO, 99+ %, 40 nm, US Research Nanomaterials, Inc), Silicon Nitride powder ($Si_3N_4$, ≥ 98.5 %, < 50 nm, Sigma-Aldrich, Inc), Silicon Dioxide ($SiO_2$, 99+ %, 20 – 30nm, US Research Nanomaterials, Inc), Iron (II, III) oxide powder ($Fe_3O_4$, 95%, < 5 µm, Sigma-Aldrich, Inc), Molybdenum Oxide ($MoO_3$, 99.9 %, 6 µm and 13-80 nm, US Research Nanomaterials, Inc)

Binder: Polycaprolactone (PCL, average Mn 80,000, Sigma-Aldrich, Inc), Polybutylene adipate terephthalate (PBAT, S-EnPol Co., Ltd)

Solvent: Tetrahydrofuran anhydrous (THF, 99.8 %, Daejung Co., Ltd), Chloroform (CF, 99.5 %, Daejung Co., Ltd)

Humectant : Tetraglycol (TG, Bioxtra, non-ionic, Sigma-Aldrich, Inc)

Conjugated molecule : Guanine (G, 98 %, Sigma-Aldrich, Inc), Indigo (IND, Dye content 95 %, synthetic, Sigma-Aldrich, Inc), Brilliant Yellow (BY, Dye content ≥ 50 %, Sigma-Aldrich, Inc)

Reagents & Buffer solution : Ultra-pure water (EXL, water purification system, 18.2 MΩ·cm @25℃), Acetic acid (glacial, 99.5%, Sigma-Aldrich, Inc), pH buffer solution (pH 4.00 ± 0.02 @25℃, pH 10 ± 0.02 @25℃, Samchun pure chemical Co., Ltd), PBS 1X solution (pH 7.4 ± 0.1, Sterile-filtered, Samchun pure chemical Co., Ltd), Agarose (Higel-agarose ClearTM, biotechnology grade, E&S Bio Electronics Company, Inc), Sodium chloride (Daejung. Inc), Glucose oxidase from Aspergillus niger (Type X-S, lyophilized powder, 100,000 – 250,000 units/g solid, Sigma-Aldrich, Inc), Ferrocene (98%, Sigma-Aldrich, Inc), D-(+)-Glucose (≥99.5% (GC), Sigma-Aldrich, Inc), Lipase from porcine pancreas (Type II, ≥125 units/mg protein, lyophilized, Sigma-Aldrich, Inc)

*Calculation of volume fraction of fillers in inks*

In the scenario where ink comprised solely of filler, binder, and solvent, the volume fraction (ρ) of filler was determined using the formula presented in equation (1) during conductivity measurements. When ink includes filler, binder, solvent, humectant, and additives, the volume fraction was computed as per equation (2) for rheological analysis. For conductivity assessments in this latter case, equation (3) was utilized to calculate the volume fraction. Here, V represents volume, M denotes mass, and d signifies material density.

$$\rho_{filler} = \frac{V_{filler}}{V_{filler} + V_{binder}} = \frac{\frac{M_{filler}}{d_{filler}}}{\frac{M_{filler}}{d_{filler}} + \frac{M_{binder}}{d_{binder}}}, \quad (1)$$

$$\rho_{filler} = \frac{V_{filler}}{V_{filler} + V_{binder} + V_{additive} + V_{humectant} + V_{solvent}} = \frac{\frac{M_{filler}}{d_{filler}}}{\frac{M_{filler}}{d_{filler}} + \frac{M_{binder}}{d_{binder}} + \frac{M_{additive}}{d_{additive}} + V_{humectant} + V_{solvent}}, \quad (2),$$

$$\rho_{filler} = \frac{V_{filler}}{V_{filler} + V_{binder} + V_{additive} + V_{humectant}} = \frac{\frac{M_{filler}}{d_{filler}}}{\frac{M_{filler}}{d_{filler}} + \frac{M_{binder}}{d_{binder}} + \frac{M_{additive}}{d_{additive}} + V_{humectant}}, \quad (3)$$

*Characterization of electrical properties of inks*

Conductive and semiconducting inks were screen-printed via home-made stencil mask on polyimide (PI) substrate or extruded via a nozzle into a mold on the glass substrate, and the geometrical parameter was measured by precision Vernier Calipers after 10 min drying. LCR meter (4100, Wayne Kerr Co., Ltd) and Digital multimeter (DT4282, HIOKI) were used to measure the resistance of the traces, and conductivity was calculated as follows.

$$\sigma = \frac{1}{\rho} = \frac{l}{wtR}$$

Terms for σ is conductivity, ρ is resistivity, R is resistance, l is the length of the resistor, w is width of the resistor, and t is the thickness of resistor. The dielectric film formed by solvent casted inks was coated with metal (Pt or Mg) on the double side of the film. Thin film deposition on nanocomposite film was performed to eliminate the gap between a metal layer and nanocomposite film which works as parasitic capacitance interfering with capacitance measurement. A digital-multi meter (DT4282, HIOKI) was used to measure the DC capacitance of the film.

$$\varepsilon_{DC} = \frac{Cd}{A}$$

Terms for ε is the dielectric constant, C is capacitance, d is the length between two metal plates which we measure as film thickness, and A is an area of the metal plate. AC capacitance of 3D-printed capacitor was characterized with an LCR meter (4100, Wayne Kerr Co., Ltd).

*Characterization of the electrochemical sintering process of the inks*

The resistance of the printed conducting trace was measured by immersion time in a 10% acetic acid solution. Probing sites for two terminals were on the surface of the printed conducting line (**Figure S1**). X-ray Diffractometer (Xpert Pro (HR-XRD), PANalytical)) analysis was performed to verify the formation of the acetate passivation layer in printed Zn trace. Scanning electron microscope (Merlin Compact (FE-SEM), ZEISS) was used to verify the existence of porosity of inks for acetic acids to be permeated through within reactive diffusion model and to verify electrochemical sintering of the interface at particles.

*Characterization of rheological properties of inks*

Rheometer (MCR 702e MultiDrive, AntonPaar) was used to characterize the rheological behavior of possible candidates of inks for 3D-printing. Inks were placed at a cavity between two plates (gap = 0.5 mm), for sweep mode, viscosity was measured within the shear rate range of 0.1 to 100 $s^{-1}$ in 25 °C and for amplitude mode, storage and loss shear modulus was measured within shear strain range of 0.01 to 10 % at angular frequency 10 rad/s. Oil was introduced at the perimeter of the top plate after ink load to prevent solvent vaporization while measurement. Change in storage modulus over drying time was measured for 15 minutes without oil at the top plate for optimization of humectant/solvent ratio.

*Formulation of biodegradable 3D-printable electronic inks*

To prepare the binder solution, start by placing 30 ml of the appropriate solvent, either THF or Chloroform, into a glass vial. Add a magnetic bar to the solvent and set the stirring speed to around 250 RPM. Next, weigh the required amount of binder pellets to achieve a concentration of 0.15 g/ml for PCL in THF or 0.2 g/ml for PBAT in Chloroform. Rapidly add the binder pellets to the stirring solvent. Seal the vial with Teflon tape and allow the mixture to homogenize. For the ink blend, measure the necessary amount of fillers and place them in a 35 ml container. Add 0.5 ml of TG to the same container. Using a norm-ject syringe, introduce 1 ml of the previously prepared binder solution into the container. Quickly close the container and seal it with Teflon tape. Weigh the total mass of the container and adjust the planetary centrifugal mixer (ARM-310, Thinky mixer) to 2000 RPM for 2.5 minutes. Finally, to load the ink, transfer it into a 3 ml barrel (EFD Norsdon) and pack it sequentially with a white piston and then an orange piston. Eject a small amount of ink and rapidly close the tip and end with a Teflon-sealed cap. To enable various printing options, replace the cap on the barrel's tip with different needles or nozzles as needed.

*Design and 3D-printing of multiple biodegradable electronic inks*

Modeling was performed with Autodesk Fusion 360, sliced with Repetier-Host, and G-code was generated for the customized components integration in complex structure. The voxel size was set at a constant value of 0.4 x 0.4 x 0.4 mm to achieve homogeneity in construction within the XYZ spatial domain. G-code was manipulated to generate a path for each material with a Hamiltonian path to reduce printing time. Then G-code was transferred to the 3D-printer (BIOX, Cellink, Inc), equipped with multi nozzles capable of printing three materials in a single layer. In instances involving the utilization of more than four distinct materials or the incorporation of post-treatment procedures between consecutive printing steps, an initial reference point was established proximal to an object on the substrate. Subsequent calibration was performed using the established standard protocol with the reference point at each sequential step. When components were printed and assembled individually, Zn-PCL ink with rapid localized electrochemical sintering was used for conductive adhesion, and PCL/THF 0.15 g/ml or PBAT/Chloroform 0.2 g/ml was used for structural adhesion.

*Characterization of degradation properties and encapsulation performances.*

Printed inks on glass are placed in a petri dish containing PBS inside and incubated in an oven at 37 °C by several time intervals. SEM image was used to observe the morphological change of fillers and binders on the surface of the printed inks by immersion time in the PBS solution. The printed encapsulation layer was examined by measuring the resistance of encapsulated electrochemically sintered Zn-PCL trace which is connected to Cu wire via silver epoxy. The time for doubling the resistance of printed Zn-PCL trace compared to initial one could be an empirical evaluation standard for the encapsulation layer.

*Characterization of mechanical properties of inks*

The inks were printed with 0.5–1 mm thickness, 5–6 mm width, and 30 mm length, and strain-stress curves were obtained by the dynamic mechanical analyzer (DMA Q800, TA Instruments, USA). Furthermore, a uniaxial tensile tester (customized, Jueuntech, Korea) and mul were used to test the stability of resistance of the Mo-PBAT trace within diverse curvatures of the printed trace.

*Measurement of energy level of inks*

Kelvin probe in Atomic Force Microscope (NX-10, Park systems) was used to measure the work function of conducting inks (Zn, Fe, Mo, and W). UPS (ultraviolet photoelectron spectroscopy) by electron spectroscopy for chemical analysis II (AXIS SUPRA, Kratos, UK) was used to measure work function and valence band level of semiconducting inks (ZnO). UV-VIS spectrometer (Microplate Spectrophotometer, Epoch 2, Bio Tek Instruments) was used to measure the direct band gap of semiconducting inks and conjugated molecules.

*Measurement of electrical performance of passive/active components*

Passive components (resistor, capacitor, inductor) were characterized by LCR meter and DMM. Active components (diode, transistor) were characterized by a semiconductor analyzer with a probe station (4200A-SCS Parameter Analyser, Keithley) by measuring two-terminal and three-terminal I-V curves. The mobility of transistor was determined as following[80]: $\mu_{Sat} = \frac{2Lg_m}{C_i W}$, where $g_m = \left(\frac{d\sqrt{I_D}}{dV_G}\right)^2$ is the transconductance; W and L is the width and length of the channel; $C_i$ is the capacitance of the dielectric layer.[81]

*Characterization of wireless system*

RF behavior of printed antenna and wireless stimulator/sensor was characterized by vector network analyzers (TTR500 Series, 100 kHz to 6 GHz, Tektronix Co., Ltd), which measured S-parameters by frequency. Wireless powering through the external device was performed by a combination of Arbitrary Function generator (AFG 31000 SERIES, Tektronix Co., Ltd) and RF amplifier (210 L, E&I Co., Ltd for small animal; High-speed bipolar amplifier BA4850, DC to 50 MHz, 8 W max, output voltage ±20 V, output current ±1 A, NF corporation Co., Ltd for large animal) with adjusted transmission frequencies and amplitudes in continuous / burst mode (interval, cycle). The wireless operation of the device was examined with a Digital Oscilloscope (TBS 1052B, 50 MHz, 1 GS/s, Tektronix Co., Ltd) with 1 kΩ loads.

*Characterization of the sensing system*

The temperature sensor was characterized with resistance measurement by LCR meter (frequency 20 Hz, 10 mV) with the sample placed on temperature varying print bed. The temperature was monitored with using an infrared camera (LIR A600-Series, FLIR, Sweden). For the pressure sensor, AC capacitance was measured using LCR meter (frequency 1 MHz, 10 mV) under an external load applied on the upper electrode, and pressure was calculated from division of load with the initial electrode area. UV sensor was characterized with conductance measurement by LCR meter (frequency 20 Hz, 10 mV) under distanced UV lamp (365 nm, 15 W). Shape-shifting photodetector array was heated with a heat gun (~80 °C) and the resistance change was characterized by digital multimeter (NI USB, 4065, National Instruments, USA). pH sensor was characterized with open circuit potential measurement in two electrode system by potentiostat (DY2100). The cell consists of a working electrode (pH sensor), reference electrode (Ag/AgCl), and pH standard solution. The glucose sensor was characterized by response current measurement in three electrode system by the potentiostat. Cell consists of working electrode (glucose sensor) with -0.02 V to reference electrode, reference electrode (Ag/AgCl), counter electrode (Pt), and glucose/PBS solution.

*Experimental setup for the wireless stimulation*

The wireless power transfer system is configured as shown in **Figure S21B**. In this set-up, the external power supply is connected to a function generator (AFG 31000 SERIES, Tektronix Co., Ltd), which can generate controlled electromagnetic waves in a pulsed form. The function generator is further connected to an RF amplifier (210 L, E&I Co., Ltd) for small animal applications, and a High-speed bipolar amplifier BA4850 (DC to 50 MHz, 8 W max, output voltage ±20 V, output current ±1 A, NF corporation Co., Ltd) for large animal applications. Additionally, the RF amplifier is connected to a coil (diameter ~15 cm, 6 turns), which facilitates the transmission of electromagnetic waves. This coil is aligned with the magnetic coupling of the antenna of wireless stimulator to maximize the efficiency of transmission. The wireless stimulator consists of an antenna with a resonance frequency of 10 – 25 MHz and a Schottky diode. The inductor used in the antenna has a resistance of 150 – 600 Ω and an inductance of approximately 800 – 980 nH at 1 MHz. For large animal stimulation, the parameters were set to an amplitude of 1.1 Vpp, a burst mode frequency of 20 Hz, and a pulse width of 200 μs. For therapeutic stimulation, the following conditions were used: an amplitude of 800 mVpp, a burst mode frequency of 20 Hz, and a pulse width of 200 μs for 1 hour.

*In vivo stimulation in small animal model (SD rat model)*

Sprague Dawley rats (male, 14-weeks old, weighing approximately 400 – 420 g) were used in the experiments. All animal care and surgical procedures were performed in compliance with the regulations of the Dankook University Animal Ethics Committee (approval number DKU-22-063). All experimental animals were individually housed in cages with sufficient food and water and maintained at a temperature of 22 - 24 °C and a humidity of 45 - 50 % in a specific pathogen-free facility. The complete sciatic nerve injury model was created in a total of 12 rats and divided into three groups: control group, non-resorbable stimulator-based stimulation group, and 3D-printed bioresorbable stimulator-based stimulation group. For the surgical procedure, anesthesia was induced with 5 % isoflurane in a mixture of oxygen and nitrogen (1:3). The hair in the left thigh area was shaved with a razor and the incision site was sterilized with povidone-iodine. The skin and subcutaneous layer were incised with scissors and the muscle layer was separated along the muscle bundle to expose the sciatic nerve. The surgical site was located at the proximal 15 mm segment of the sciatic nerve, where the tibial and common peroneal nerves branch out from the sciatic nerve, and extended to the hip joint. The sciatic nerve was completely cut using microscissors and 3D-printed bioresorbable stimulator was interfaced via biodegradable Mo wires (laser patterned Mo foil; 0.0125 mm, 99.9%, Goodfellow Cambridge Ltd.) on the proximal segment from the injury site (**Figure S20**). The nerve was sutured using 10-0 Nylon, and the surrounding muscles and skin were sutured using 6-0 ethilon after nerve repair. In the control group, a non-resorbable stimulator was implanted into the same surgical site, and the nerve was sutured using the same method as the experimental group. After surgery, the rats were allowed free movement without any special fixation.

*In vivo stimulation in large animal model (Canine)*

6-month-old female beagle was used in the experiment under the following breeding conditions: temperature was maintained at 23±2 °C, humidity was maintained at 50 -60 %, and ventilation was maintained at 10 - 20 times per hour. All animal care and surgical procedures were performed in compliance with the regulations of the KBIO Institutional Animal Care and Use Committee (approval number KBIO-IACUC-2021-275). The animals were pre-anesthetized with Zoletil (5 mg/kg) and xylazine (2 mg/kg) and intubated for inhalation anesthesia with isoflurane (3 %) while monitoring their anesthetic state using an ECG, $SpO_2$, and $EtCO_2$ monitoring system. Prior to the skin incision, 2 % lidocaine was administered at the incision site for local anesthesia. The skin, subcutaneous tissue, and muscles above the sciatic nerve were sequentially incised to expose the sciatic nerve. The sciatic nerve was cut approximately 2 cm from the distal end of the femoral joint, and a 3D-printed bioresorbable electrical stimulator was directly threaded by the proximal segment and the incision was sutured (**Figure S21A**). The operation of the 3D-printed electrical stimulator was verified using CMAP and CNAP (**Figure S22, Figure S23**). The mechanical stability of the stimulator in vivo was confirmed by C-arm imaging after its implantation.

*Electrophysiology*

The rats were positioned on their sides and the active electrode was attached to the rats' biceps femoris muscle, with the reference electrode attached to the Achilles tendon and the ground electrode attached to the tail. CMAP was measured with EMG measurement equipment (IX-RA-834, iWorx, Co., Ltd). In vivo stimulation in large animal was verified by CNAP signals obtained in the proximal segment of the sciatic nerve. The active electrode and reference electrode were interfaced to the proximal region from a 3D-printed bioresorbable stimulator that is already embracing the sciatic nerve and a ground electrode attached to distal musculature. CNAP was measured with precise measurement equipment (Neuropack S1, MEB-9400, Nihon Kohden Co., Ltd).

*Ultrasound-mediated degradation of stimulator in small animal model (SD rat)*

Ultrasound was triggered with ultrasound irradiator (MV-100, Mirae Ultrasonic Co.,Ltd, Korea) after the stimulator was implanted to the SD rat model. Ultrasound was applied after applying ultrasonic gel (GREENSONIC, Greenpharm, Korea) to prevent attenuation of the ultrasound. The intensity of ultrasound was 3.5 W/cm$^2$, with frequency of 20568 Hz and ultrasound was irradiated for 30 minutes. After irradiation, the image of the device was obtained from X-ray micro-CT (Skyscan 1176, Bruker).

## Data availability

All data are available in the main text or the supplementary materials.

## Acknowledgements

This research was supported by National R&D Program through the National Research Foundation of Korea (NRF) funded by Ministry of Science and ICT (2022R1C1C1008513 and 2022M3H4A1A04096393). Authors acknowledge the use and support of Anton Paar Korea for rheological characterization.

## Author contributions

Conceptualization: JYL, JJ, JKH, SKK

Methodology: JYL, SKK

Investigation: JYL, JJ, JHP, SHK, YSP, MSC, JH, KSK

Visualization: JYL, JJ, JHL, SGC, SYP, YSK, YNK, SML, MKC, JMM, JWK, SKS, JK, JK, JK, WBK, KSL, JKH, KSK

Funding acquisition: SKK

Project administration: JKH, SKK

Supervision: JKH, SKK

Writing – original draft: JYL, JJ, JKH, SKK

Writing – review & editing: JYL, JJ, JHP, SHK, JKH, SKK

## Competing interests

Authors declare that they have no competing interests. Two patents (KR 10-2439010 and KR 10-2490474) and two patent applications (PCT application no. PCT/KR2020/018179 and application no. US 18/015956) have been filed related to the topic covered in this publication.

## Additional Information

Supplementary information is available for this paper. Supplementary information includes Supplementary Notes 1-2, Supplementary Figures 1-25, Supplementary Table 1,2, Supplementary Videos 1-6, and Supplementary References.

**Correspondence and requests for materials** should be addressed to Jung Keun Hyun and Seung-Kyun Kang.